\documentclass[final]{l4dc2026}
\usepackage{times}
\usepackage{xurl}
 
\definecolor{NCSUred}{RGB}{153, 0, 0}
\definecolor{NCSUgreen}{RGB}{0, 132, 115}
\definecolor{NCSUblue}{RGB}{65, 86, 161}
\definecolor{NCSUorange}{RGB}{209, 73, 5}

\newcommand{\mc}[1]{\mathcal{#1}}
\newcommand{\ms}[1]{\mathsf{#1}}
\newcommand{\mr}[1]{\mathrm{#1}}
\newcommand{\mb}[1]{\mathbf{#1}}
\newcommand{\mbb}[1]{\mathbb{#1}}
\newcommand{\mR}{\mathbb{R}}
\newcommand{\mC}{\mathbb{C}}

\newcommand{\xD}[1]{\mr{d} #1}

\newcommand{\bra}[1]{\left( #1 \right)}
\newcommand{\Bra}[1]{\left[ #1 \right]}
\newcommand{\BRA}[1]{\left\{ #1 \right\}}
\newcommand{\norm}[1]{\left\| #1 \right\|}
\newcommand{\ip}[2]{\langle #1, \, #2 \rangle}

\usepackage{xcolor}
\usepackage{subcaption}

\long\def\comment#1{}

\newcommand{\be}{\begin{equation}}
\newcommand{\ee}{\end{equation}}
 
\newfont{\bbb}{msbm10 scaled 700}

\newfont{\bb}{msbm10 scaled 1100}

\newcommand{\<}{\left\langle}
\renewcommand{\>}{\right\rangle}

\usepackage{booktabs}
\usepackage{epstopdf}
\usepackage{multirow}

\title[Koopman Operator for Stability Analysis]{Koopman Operator for Stability Analysis: \\
Theory with a Linear--Radial Product Reproducing Kernel}
\coltauthor{\Name{Wentao Tang} \Email{wtang23@ncsu.edu} \and \Name{Xiuzhen Ye} \Email{xye7@ncsu.edu} \\
\addr Department of Chemical and Biomolecular Engineering, North Carolina State University}

\begin{document}

\maketitle

\begin{abstract}
    Koopman operator, as a fully linear representation of nonlinear dynamical systems, if \emph{well-defined} on a reproducing kernel Hilbert space (RKHS), can be efficiently learned from data. 
    For stability analysis and control-related problems, it is desired that the defining RKHS of the Koopman operator should account for both the stability of an equilibrium point (as a \emph{local} property) and the regularity of the dynamics on the state space (as a \emph{global} property). 
    To this end, we show that by using the \emph{product kernel} formed by the linear kernel and a Sobolev radial kernel, the resulting RKHS is invariant under the action of Koopman operator (under certain smoothness conditions). 
    Furthermore, when the equilibrium is asymptotically stable, the spectrum of Koopman operator is provably confined inside the unit circle, and escapes from the unit disk upon bifurcation. Thus, the learned Koopman operator with provable probabilistic error bound provides a \emph{stability certificate}. 
    In addition to numerical verification, we further discuss how such a fundamental \emph{spectrum--stability relation} would be useful for Koopman-based control. 
\end{abstract}


\section{Introduction}
The pursuit of a generic linearization approach to efficiently solve nonlinear systems and control problems gives rise to the popularity of Koopman operators (composition operators) \citep{mauroy2020koopman, brunton2022modern}. 
For a discrete-time dynamics with transition map $f$ on the state space $\mbb{X}$, the Koopman operator $A$, which is a linear operator, maps any state-dependent function $g \in \mc{G}$ to its composition with the dynamics: $Ag = g\circ f$. 
For $A$ to be well-defined, the function space $\mc{G}$ must at least be \emph{invariant} under the composition. To use such an operator, we need it to be \emph{learnable} from data with provably bounded errors. 
Moreover, it is desired that the Koopman operator can recover system properties that are of interest to control---most importantly, the \emph{Koopman spectrum} should reflect the \emph{stability} of an equilibrium point. 
While such an objective is natural to control theorists, the \emph{spectrum--stability relation of the learned Koopman operator} remains underexplored, and we aim to address this issue with new theoretical constructions in this paper.

\paragraph{Function space defining the Koopman operator.}
For a measure-preserving dynamics with invariant measure $\pi$, the Koopman operator is well-defined on $\mc{G} = L^2_\pi$ \citep{arbabi2017ergodic}. 
In this setting, a suitable approach to learn the Koopman operator is extended dynamic mode decomposition (EDMD) \citep{williams2015data}, where the state data is lifted by a dictionary of nonlinear functions (which should form a basis of $L^2_\pi$ \citep{korda2018convergence}), and a linear dynamics is learned in the lifted dimension. 
% Specifically, if choosing the dictionary by a truncated basis of $L^2_\pi$ and sampling data under $\pi$, EDMD converges provably to the true Koopman operator. 
For more general dynamics, \cite{mezic2020spectrum} defined the Koopman operator on $L^2_\pi$ to capture on-attractor measure-preserving dynamics and a Segal--Bargmann space to capture off-attractor attractive dynamics. 
However, for the EDMD implementation, it tends to be impractical to assume prerequisite knowledge about the invariant measure, basis functions on a general domain, or location of the attractor set in the state space. 
 
\par To learn the Koopman operator, it is convenient to adopt a RKHS formulation and use a kernel-based regression algorithm \citep{kevrekidis2016kernel}. If a RKHS is not invariant, the operator learned can only be the Koopman operator composed with some injection and projection operators \citep{klus2020eigendecompositions, kostic2022learning, philipp2024error}. 
Recently, \cite{kohne2025error} found that the Koopman operator is well-defined on a Sobolev-type Hilbert space under certain smoothness and non-degeneracy conditions on $f$. Such a Sobolev--Hilbert space (with a sufficient smoothness index) is in fact an RKHS \citep{wendland2004scattered}. 
Hence, an RKHS- (and hence $L^\infty$-) error bound of single-step state prediction with learned Koopman operator was deduced \citep{kohne2025error}.

\vspace{-0.15em}
\paragraph{Stability in Koopman operator modeling.} 
Typically, for the Koopman operator learned from snapshot sample, the error bound is established on its one-time-step state prediction \citep{philipp2024error, kohne2025error}; long-horizon prediction capability can be improved by using long-horizon data and trajectory-based kernels \citep{bevanda2023koopman}. Yet, from a control point-of-view, the \emph{stability of an equilibrium point}, as a qualitative property, is of special importance. 

If an equilibrium point is \emph{a priori} known to be stable, \cite{bevanda2022diffeomorphically, fan2022learning} proposed to force the learned EDMD matrix to be stable. 
\cite{breiten2023approximability} and \cite{tang2024koopman} used weight factors to create weighted function spaces ($L^p$, continuous, and RKHS) to define the Koopman operator. Specifically, the weight factor is chosen in concert with the convergence rate to the equilibrium point, so as to ``modulate'' the local regularity or singularity of functions in scope. 
As such, \cite{tang2024koopman} verifies that the Koopman operator is contractive and usable for obtaining a Lyapunov/Zubov function certificate. In a recent preprint,  \cite{breiten2025unifying} showed the well-definedness of the value function under optimal control. 
In this paper, we do not assume such prior knowledge of stability; instead, we only assume the presence (location) of an equilibrium point (say, the origin) and let the learned Koopman operator disclose its stability property. In other words, we aim to allow \emph{stability analysis} based on the learned Koopman operator.

\vspace{-0.15em}
\paragraph{Koopman-based control.} A benefit of this work is to bring us closer to a \emph{theoretical foundation of Koopman-based control}. Let us briefly remark on this point, with more discussions later in Section \ref{sec:discussions}. 
Indeed, there was such optimism that with a linear operator, nonlinear control problems could be reduced formally to linear ones \citep{proctor2018generalizing, korda2018linear}. 
Realizing that full linearity is untruthful, ``lifting'' into a bilinear system has been considered \citep{goswami2021bilinearization, bruder2021advantages}. Theoretically, operator representation for such bilinear systems can be constructed \citep{iacob2024koopman, bevanda2024nonparametric, tang2025koopman}. 
However, unnecessary complications often arise from bilinearity, e.g., a state-feedback law guarantees only local convergence \citep{strasser2024koopman}, while stability for Koopman-based (open-loop) MPC can only be {\it ad hoc} verified (instead of directly synthesized) \citep{bold2025kernel}. 

\par At this point, a solid foundation of Koopman-based control is still lacking \citep{berberich2024overview, haseli2025two}. A key reason is the \emph{elusive relation between the Koopman spectrum and stability}. 
If a RKHS is specified by a radial kernel (as in \cite{kohne2025error}), then $\|A\|\geq1$ always hold, regardless of stability. 
While the Segal--Bargmann space in \cite{mezic2020spectrum} is an RKHS that gives a clear \emph{spectrum--stability relation}, its kernel is difficult to calculate from data due to the use of an implicit homeomorphism that converts $f$ to a complexified linear dynamics.

\paragraph{Contributions of this work.} First, we propose a new defining function space for the Koopman operator. Specifically, we use a product kernel formed by the linear kernel and the Sobolev radial kernel, the former of which accounting for the existence of equilibrium point and the latter accounting for the regularity of $f$.  
Second, by analyzing the structure of the function members of this RKHS and using diffeomorphism arguments, we find that the stability of the equilibrium point is reflected in the spectrum of the Koopman operator on the RKHS. 
Third, building on the error bound of operator learning on this new RKHS, we establish that the finite-rank operator estimated from data can be used for stability certification and bifurcation detection, which we confirm with a numerical experiment. 
Finally, we discuss how this work can be potentially useful towards a Koopman-based control theory. For brevity of presentation in the main text, all proofs are provided in Appendix \ref{app:proofs}.

\section{Preliminaries}\label{sec:preliminaries}
We consider a discrete-time dynamical system 
\begin{equation}\label{eq:system}
    x_{t+1} = f(x_t), \enspace f: \mbb{X}\rightarrow \mbb{X} \subset \mR^{d}. 
\end{equation}
For any state-dependent function $g: \mbb{X}\rightarrow \mR$, we refer to its composition with the dynamics $f$ as $g\circ f$, which maps any $x\in \mbb{X}$ to $g(f(x))$. The Koopman operator $A$ of system \eqref{eq:system} is defined as the following linear mapping: 
\begin{equation}\label{eq:Koopman}
    A: \mc{G}\rightarrow \mc{G}, g \mapsto g \circ f, 
\end{equation}
on function space $\mc{G}$, which must be invariant: $A\mc{G} = \{Ag: g\in \mc{G}\} \subset \mc{G}$. 
Except for very special examples, the invariance of $\mc{G}$ requires it to be infinite-dimensional. Despite this, with regularity conditions on $f$, we may choose $\mc{G}$ to be just ``as large as necessary''. 

\subsection{Koopman operator on Sobolev--Hilbert space} 
We use the following conclusion in \cite{kohne2025error}, which defines the Koopman operator on a Sobolev--Hilbert space $W^{s, 2}(\mbb{X})$. By this notation, we refer to the space of functions with all generalized derivatives up to the $s$-th order being square-integrable on $\mbb{X}$. 
The invariance of $W^{s,2}(\mbb{X})$ requires that the dynamics $f$ be $C^s$ (have derivatives up to the $s$-th order). In applications where $f$ is governed by physical laws in analytical equations, such a smoothness condition is mild. 
\begin{lemma}\label{lem:smoothness}
    Suppose that $\mbb{X}\subset \mR^d$ is compact. If $f\in C^s(\mbb{X})$ and is non-degenerate in the sense of $\inf_{x\in \mbb{X}} \lvert \mr{det}~\mr{D}f(x) \rvert > 0$, then $A$ is a bounded linear operator on $W^{s,2}(\mbb{X})$. 
\end{lemma}

The smoothness parameter $s$ here can take fractional values; the corresponding $W^{s,2}$ is known as the Sobolev--Slobodeckij spaces. The lemma above still holds, if $C^s(\mbb{X})$ with a fractional $s$ is interpreted as a H{\"{o}}lder function space \citep{adams2003sobolev}. We know from the Sobolev embedding theorems that when $s>d/2$, $W^{s,2}(\mbb{X})$ can be compactly embedded into the continuous function space $C(\mbb{X})$. In fact, it is possible to make $W^{s,2}(\mbb{X})$ coincide with an RKHS.

\subsection{Sobolev--Hilbert space as an RKHS} 
We refer to a continuous and symmetric bivariate function $\kappa: \mbb{X}\times \mbb{X}\rightarrow \mR$ as a \emph{Mercer kernel} if for any $\{x^{(i)}\}_{i=1}^n \subset \mbb{X}$, the matrix $G_\kappa = \Bra{ \kappa\bra{x^{(i)}, x^{(j)}} }_{i,j=1}^n$ is positive semidefinite. 
The closed span of all kernel functions $\kappa(x, \cdot)$ with $x\in \mbb{X}$, endowed with inner product: $\ip{\kappa(x,\cdot)}{\kappa(x',\cdot)} = \kappa(x,x')$ and hence norm: $\|\kappa(x, \cdot)\| = \kappa(x,x)^{1/2}$, is called the RKHS specified by kernel $\kappa$ \citep{steinwart2008support}. We denote this RKHS as $H_\kappa(\mbb{X}) = H_\kappa$. The following conclusion from \cite{wendland2004scattered} states the equivalence of Sobolev--Hilbert spaces with RKHSs. 
\begin{lemma}\label{lem:Sobolev-RKHS}
    Suppose that $\kappa$ is a \emph{radial kernel}, i.e., $\kappa(x, x') = \rho(|x-x'|)$ for some $\rho: \mR_+\rightarrow \mR_+$, and that the Fourier transform of $\rho$, $\hat{\rho}(\xi)$, satisfies $c_1(1+|\xi|^2)^{-s}\leq |\hat{\rho}(\xi)| \leq c_2(1+|\xi|^2)^{-s}$ for some constants $c_2\geq c_1>0$. In addition, assume that $\mbb{X}$ is a region with Lipschitz boundary. Then $W^{s,2}(\mbb{X})$ coincides with $H_\kappa(\mbb{X})$, with equivalent norms. 
\end{lemma}

\par \cite{wendland2004scattered} provided the following kernel to satisfy the conditions in Lemma \ref{lem:Sobolev-RKHS}. 
\begin{lemma}\label{lemma:Wendland}
    For any $k\in \mbb{N}$ (including $0$), the following function $\rho: \mR_+\rightarrow \mR_+$ satisfies $c_1(1+|\xi|^2)^{-s/2}\leq |\hat{\rho}(\xi)| \leq c_2(1+|\xi|^2)^{-s/2}$ for some constants $c_2\geq c_1>0$ and $s=\frac{d}{2}+k$: 
    \begin{equation}\label{eq:Wendland.kernel}
        \textstyle \rho = I^k \varrho_{\lfloor d/2+k+1 \rfloor}, \text{ where } \varrho_\ell(r) = \max\{1-r, 0\}^\ell \text{ and } I\phi(r) = \int_r^\infty r'\phi(r')dr'. 
    \end{equation}
    The resulting radial kernel $\kappa$ is hereforth denoted as ``$\kappa^=$'' (without explicitly stating the index $k$).   
\end{lemma} 
\begin{corollary}\label{cor:Wendland}
    Suppose that (i) $\mbb{X}$ is a bounded region with Lipschitz boundary, (ii) $f\in C^s(\mbb{X})$ with $s=\frac{d}{2}+k$, $k\in \mbb{N}$, and (iii) $\inf_{x\in \mbb{X}} \lvert \mr{D}f(x) \rvert > 0$. Then $A$ is a bounded linear operator on $H_{\kappa^=}(\mbb{X})$. 
\end{corollary}

\par Importantly, on the RKHS, the reproducing property hold: $\ip{g}{\kappa(x, \cdot)} = g(x)$ for all $g\in H_\kappa(\mbb{X})$ and $x\in \mbb{X}$. Thus, when $A: H_\kappa\rightarrow H_\kappa$ is well-defined, we can also define the adjoint operator $A^\ast: H_\kappa\rightarrow H_\kappa$, called the \emph{Perron-Frobenius operator} \citep{klus2020eigendecompositions}. 
On an RKHS $H_\kappa$, it is easily verified that $A^\ast$ is the operator pushing forward the kernel function: $A^\ast \kappa(x, \cdot) = \kappa(f(x), \cdot)$, $\forall x\in \mbb{X}$. If $A$ is bounded, $A^\ast$ is also bounded. The benefit of considering the Perron-Frobenius operator is to learn it from data according to the foregoing property. 

\subsection{Koopman operator learning by reduced rank regression}\label{sec_Koopman operator learning by kernel EDMD}
Suppose that the Koopman operator $A$ is well-defined on $H_\kappa$. Provided an independent and identically distributed sample of snapshots $\{x^{(i)}, y^{(i)}\}_{i=1}^n$ where $y^{(i)} = f(x^{(i)})$ for each $1\leq i\leq n$, an operator $\hat{A}^\ast$ from $H_\kappa$ to $H_\kappa$, as an approximated Perron-Frobenius operator, can be learned to match $\hat{A}^\ast \kappa(x^{(i)}, \cdot)$ with $\kappa(y^{(i)}, \cdot)$ on the sample. 
The problem is formulated as a convex optimization one over the space of Hilbert--Schmidt operators, which, for the sake of controlling the generalization error, is restricted to operators with rank not exceeding $r$ \citep{kostic2022learning}: 
\begin{equation}\label{eq:kEDMD}
\textstyle
    \min_{\hat{A}^\ast \in \mr{HS}(H_\kappa):\, \mr{rank}~\hat{A}^\ast = r} \enspace \frac{1}{n}\sum_{i=1}^n \norm{\hat{A}^\ast \kappa\bra{x^{(i)}, \cdot} - \kappa\bra{y^{(i)}, \cdot}}_{H_\kappa}^2 + \beta\|\hat{A}^\ast\|_{\mr{HS}}^2.   
\end{equation}

\par Since the space of Hilbert--Schmidt operators is a Hilbert space itself, the representer theorem \citep{smola1998learning} applies, which reduces the optimization problem onto a space of finite-rank operators: $\sum_{i=1}^n \sum_{j=1}^n \theta_{ij} \kappa(y^{(i)}, \cdot)\times \kappa(x^{(j)}, \cdot)$, where $\kappa(y^{(i)}, \cdot)\times \kappa(x^{(j)}, \cdot)$ is a rank-$1$ operator mapping $\kappa(x, \cdot)$ to $\kappa(x^{(j)}, x) \kappa(y^{(i)}, \cdot)$. 
We omit the detailed derivations for the final solution of the optimal solution $\hat{\Theta} = [\hat{\theta}_{ij}]_{i,j=1}^n$. 
Once $\hat{\Theta}$ is solved, the estimation of $A$ is therefore $\hat{A} = \sum_{i=1}^n \sum_{j=1}^n \hat{\theta}_{ij} \kappa(x^{(j)}, \cdot) \times \kappa(y^{(i)}, \cdot)$. The statistical error of $\hat{A}$ was analyzed in \cite{kostic2023sharp}, which we will utilize in Section \ref{sec:spectrum}. 

\par Given an initial state $x_0 = x$, the estimated operator $\hat{A}^\ast$ can be used to predict the next state. Since $\kappa(x_1, \cdot) \approx \hat{A}^\ast \kappa(x_0, \cdot) = \sum_{i,j} \hat{\theta}_{ij} \kappa(x^{(j)}, x_0) \kappa(y^{(i)}, \cdot)$, we have $x_1 \approx \sum_{i,j} \hat{\theta}_{ij} \kappa(x^{(j)}, x_0) y^{(i)}$. 
For $t$-step prediction, by recursion on $\hat{A}^\ast$, we can write 
\begin{equation}\label{eq:state.prediction}
\textstyle
    x_t \approx \sum_{i_t,j_t, \dots,i_1,j_0} \Bra{\prod_{\tau=1}^{t} \hat{\theta}_{i_\tau j_\tau} \kappa\bra{x^{(i_{\tau})}, y^{(j_{\tau-1})}} } \hat{\theta}_{i_0 j_0} \kappa\bra{x^{(j_0)}, x_0} y^{(i_t)}. 
\end{equation}
Clearly, the spectrum of the EDMD matrix representation $\hat\Theta$ plays a major role in the prediction accuracy. 
In particular, it is desired that when the equilibrium point at the origin asymptotically attracts the orbits on $\mbb{X}$, the spectrum of $\hat{\Theta}$ can be guaranteed to reside on the open unit disk (when the sample size is sufficiently large). Hence, we aim to make the RKHS be ``aware'' of the equilibrium point, for which a radial kernel that is ``indifferent'' to a special point would be unsuitable.

\section{Koopman Operator using Linear--Radial Product Kernel}\label{sec:kernel}
The ``trick'' proposed in this paper for the RKHS to be ``aware'' of the equilibrium point is to use a product kernel. 
For this, we recall the definition of the \emph{(algebraic) tensor product} of two Hilbert spaces $\mc{G}$ and $\mc{H}$ as the completed span of elementary tensors \citep{lax2002functional}:
$$ \mc{G} \otimes \mathcal{H} = \overline{\mr{span}} \{g \otimes h: g \in \mc{G}, \, h \in \mc{H}\}, $$
which is a Hilbert space endowed with the inner product for elementary tensors $\ip{g_1 \otimes h_1}{g_2 \otimes h_2} = \langle g_1,g_2\rangle_{\mc{G}} \langle h_1,h_2\rangle_{\mc{H}}$ and hence the induced norm $ \norm{\sum_{i=1}^{n} g_i \otimes h_i}_{\mc{G}\otimes \mc{H}}= 
\sqrt{\sum_{i,j=1}^{n}
\langle g_i,g_j\rangle_{\mc{G}}\,
\langle h_i,h_j\rangle_{\mc{H}}}$. 
We are interested in the case with RKHSs: $\mc{G} = H_\lambda(\mbb{X})$ and $\mc{H} = H_\mu(\mbb{Y})$, where $\lambda$ and $\mu$ are kernels on set $\mbb{X}$ and set $\mbb{Y}$, respectively. Then for any elementary tensor $g \otimes h$, we can verify that 
$$ \ip{g \otimes h}{\lambda(x, \cdot) \otimes \mu(x, \cdot)}_{{\cal G} \otimes {\cal H}} = \langle g, \lambda(x, \cdot) \rangle_{\cal G} \langle h, \mu(x, \cdot) \rangle_{\cal H} = g(x)h(y). $$
Hence we see that $\mc{G}\otimes \mc{H}$ is actually an RKHS on $\mbb{X}\times\mbb{Y}$, whose kernel function $\kappa$ is specified by $\kappa(\cdot, (x, y)) = \lambda(\cdot, x)\mu(\cdot, y)$ -- namely the product of two kernels.
In particular, if $\mbb{X} = \mbb{Y}$, and we are only interested in taking all members of $H_\lambda\otimes H_\mu$ at one point $x$ rather than a pair of points $(x,y)$, then we may equate $g\otimes h$ to $gh$. That is, for $\mc{G} = H_\lambda(\mbb{X})$ and $\mc{H} = H_\mu(\mbb{X})$, 
\begin{equation}\label{eq:tensor}
    H_\lambda(\mbb{X}) \otimes H_\mu(\mbb{X}) = \overline{\mr{span}}\BRA{g\otimes h: g\in H_\lambda(\mbb{X}), \, h\in H_\mu(\mbb{X})} = H_\kappa(\mbb{X}), \text{ where } \kappa = \lambda\mu.
\end{equation}

\subsection{Linear--radial product kernel} 
For a special attention on the equilibrium point behavior, we focus on the design of the kernel function $\kappa$ specifying the RKHS. 
We know that for a finite ($d$)-dimensional linear system, the Koopman operator is well-defined on the space of linear functions on $\mR^d$, which is equivalent to the RKHS specified by the linear kernel: $\kappa^-(x,x') = x^\top x'$, as long as the equilibrium point (assumed to be the origin) is in the interior of $\mbb{X}\subset \mR^d$. (We use superscript ``$-$'' to symbolize the word ``linear''.) That is, $ H_{\kappa^-} = \BRA{ x\mapsto c^\top x: c\in \mbb{X}}$. 
In view of \eqref{eq:tensor}, we shall put $\kappa^-$ as one of the factors in $\kappa$. 

\par However, for a nonlinear system, the Koopman operator is not well-defined on $H_{\kappa^-}$. Then, we shall have a second kernel as a factor in $\kappa$ that guarantees the invariance of the Koopman operator. 
By Corollary \ref{cor:Wendland}, assuming knowledge on the regularity of the dynamics, the corresponding Sobolev kernel in \eqref{eq:Wendland.kernel} specifies a RKHS that accommodates the Koopman operator. 
We denote by $\kappa^=$ such a radial kernel. (The symbol ``$=$'' means the translation invariance of the radial kernel.) At this point, we recall from Lemma \ref{lem:Sobolev-RKHS} that $H_{\kappa^=} \equiv W^{s,2}(\mbb{X})$. 
Therefore, we define from now:
\begin{equation}\label{eq:product.kernel}
    \kappa = \kappa^- \kappa^= \text{ and } \mc{H} = H_{\kappa}(\mbb{X}) = H_{\kappa^-}(\mbb{X}) \otimes H_{\kappa^=}(\mbb{X}). 
\end{equation}

\par We thus have the following lemma, which describes the function members in the above-defined RKHS as ``$W^{s,2}$-functions multiplied by linear functions''. Colloquially speaking, $g\in \mc{H}$ if $g$ is ``$s$-smooth in the generalized (Sobolev) sense'' and ``locally at least linear'' near the origin. 
\begin{lemma}\label{lem:member}
    Suppose that $f(0)=0$, $0\in \mr{int}(\mbb{X})$, and $f\in C^s(\mbb{X})$ with $s=\frac{d}{2}+k$ ($k\in \mbb{N}$). Let $\kappa^-$ be the linear kernel and $\kappa^=$ be the $k$-th Wendland kernel. Then $\mc{H}$ defined in \eqref{eq:product.kernel} satisfies:
    \begin{equation}\label{eq:member}
    \textstyle
        \mc{H} = \BRA{\sum_{i=1}^d e_ih_i: h_1,\dots,h_d\in W^{s,2}(\mbb{X})}.
    \end{equation}
    where $e_i(x) = x_i$, $i=1,\dots,d$, are projections onto components. Moreover, 
    $$ \textstyle \text{for } g = \sum_{i=1}^d e_ih_i \in \mc{H} \text{, we have } \norm{g}_{\mc{H}}^2 = \sum_{i=1}^d \norm{h_i}_{H_{\kappa^=}}^2. $$
\end{lemma}

\subsection{Well-definedness of Koopman operator on the RKHS} 
When defining the Koopman operator on the RKHS specified by the radial kernel $\kappa^=$, namely $W^{s,2}(\mbb{X})$, it was required that the dynamics $f$ is $C^s$ (cf. Lemma \ref{lem:smoothness}). To ensure the well-definedness with the product kernel $\kappa = \kappa^-\kappa^=$, we should adapt the requirement on $f$. For this, we define the following function class:
\begin{equation}\label{eq:F.class}
\textstyle
    \Phi^s(\mbb{X}) := \BRA{\sum_{i=1}^d e_i\phi_i: \phi_1, \dots,\phi_d\in C^s(\mbb{X}, \mR)} , 
\end{equation}
which is informally interpreted as the space of $s$-smooth functions that have zero values at the origin. 

\begin{theorem}\label{th:definedness}
    Suppose that (i) $\mbb{X}\subset\mR^d$ is compact, (ii) the component functions of the dynamics $f$: $f_1, \dots, f_d \in \Phi^s(\mbb{X})$, and (iii) $f$ is non-degenerate: $\inf_{x\in \mbb{X}} \lvert \mr{det}~\mr{D}f(x) \rvert > 0$. Then the Koopman operator $A$ is a bounded linear operator on $\mc{H}$ (as defined in \eqref{eq:product.kernel}). 
\end{theorem}

% \begin{remark}
%     The Wendland kernel in~\eqref{eq:Wendland.kernel} is such that
% $(I^{k}\mathring{\rho}_{\ell}^{\mathrm{W}})(r)
% = \int_{r}^{1} t_1
%    \int_{t_1}^{1} t_2
%    \cdots
%    \int_{t_{k-1}}^{1} t_k (1-t_k)^{\ell}\,dt_k \cdots dt_2\,dt_1$.
% Large $k$ indicates a more smooth of the resulting function. The "basis" function $\mathring{\rho}_{\ell}^{\mathrm{W}}$ is continuous at $r = 1$. 
% $\frac{d \mathring{\rho}_{\ell}^{\mathrm{W}}}{dr}$ are nonsmooth only at the endpoint $r = 1$. Therefore, increase $k$ by one leads to the increase of smoothness by $2$. 
% \end{remark}

\section{Koopman Spectrum and Stability of Equilibrium Point}\label{sec:spectrum}
Now that the Koopman operator is well-defined on the RKHS with the linear--radial product kernel, we examine the role of Koopman spectrum in stability analysis. 
It is naturally postulated that if the origin is an asymptotically stable equilibrium point attracting all orbits from $\mbb{X}$, then the Koopman spectrum is in $\mbb{D} = \{z\in \mC:|z|<1\}$. Clearly, for all $g = \sum_{i=1}^d e_ih_i \in \mc{H}$, $A^t g(x) = \sum_{i=1}^d (f^t(x))_i h_i(f^t(x)) \rightarrow \sum_{i=1}^d 0\cdot h_i(0) = 0$ as $t\rightarrow \infty$. 
On the other hand, if the origin becomes unstable, then orbits that move away from the origin shall be used to show that the Koopman spectrum is not disjoint from the complement of $\overline{\mbb{D}}$. 
% \par For the convenience of analysis (especially error analysis via the lens of \emph{perturbation theory}), we note that $T=A^\ast A$ as a \emph{self-adjoint} operator on the Hilbert space $\mc{H}$. Actually, we will consider its spectrum, which is restricted to the real line: $\sigma(T)\subset \mR$, instead of $\sigma(A)$.  

\subsection{Spectrum under stability} 
We aim to show that the spectrum of $\sigma(A)$, when the origin is asymptotically stable, must be a subset of $\mbb{D}$, as long as certain \emph{homeomorphism} conditions are satisfied. 
By assuming the existence of a homeomorphism between the dynamics $f$ and the $d$-dimensional linear system $\tilde{f}: z\mapsto Fz$ governed by the Jacobian $F = \mr{D}f(0)$, the characterization of eigenfunctions and eigenvalues Koopman operator on $L^2$- and Segal--Bargmann function spaces were dis cussed in \cite{mezic2020spectrum}. 
By a homeomorphism, we refer to $\psi: \mbb{X}\rightarrow \mbb{Z} = \psi(\mbb{X})$ such that $\psi\circ f = \tilde{f}\circ \psi$. In particular, we say that $\psi$ is a \emph{$\Phi^s$-homeomorphism} if $\psi\in (\Phi^s(\mbb{X}))^d$, $\psi$ is invertible, and $\psi^{-1}\in (\Phi^s(\mbb{Z}))^d$. 
\begin{lemma}\label{lem:homeomorphism}
    Suppose that the conditions in Theorem \ref{th:definedness} hold, and in addition, there exists a $\Phi^s$-homeomorphism $\psi$ such that $\tilde{f} = \psi\circ f\circ \psi^{-1}: z\mapsto Fz$. Let $\tilde{\mc{H}} = \BRA{ \sum_{i=1}^d e_i\tilde{h}_i: \tilde{h}_i\in W^{s,2}(\mbb{Z}) }$. Then $\tilde{A}: \tilde{\mc{H}}\rightarrow \tilde{\mc{H}}$, $\tilde{g}\mapsto \tilde{f}\circ \tilde{g}$ is a linear and bounded operator. Moreover, $\sigma(\tilde{A}) = \sigma(A)$. 
\end{lemma}

Roughly speaking, the ``homeomorphic-to-linear'' condition of Lemma \ref{lem:homeomorphism} requires that the equilibrium point at $0$ is the sole \emph{invariant structure in $\mbb{X}$}. Naturally, if another invariant set (e.g., a limit cycle) exists in $\mbb{X}$, then $\sigma(A)$ may be a reflection of attractiveness of such a structure instead. 
Following the above lemma, the following theorem states that under the above assumed conditions, $\sigma(A)$ serves as a certificate of the stability of the equilibrium point.
\begin{theorem}\label{th:spectrum}
    Assume the conditions in Theorem \ref{th:definedness} and a $\Phi^s$-homeomorphism $\psi$ such that $\tilde{f} = \psi\circ f\circ \psi^{-1}: z\mapsto Fz$. Then $\sigma(A) = \overline{\< \sigma(F) \>}$, where $\< \sigma(F) \> = \BRA{\prod_{i=1}^r \lambda_i: \lambda_i \in \sigma(F), \, r \in \mbb{N}\backslash \{0\}}$ is the semigroup generated by the set of eigenvalues of $F$. In particular, if $F$ is stable, then $\sigma(A)\subset \mbb{D}$. 
\end{theorem}

\subsection{Spectrum upon bifurcation} 
If the equilibrium point becomes unstable, it may not be guaranteed that $\sigma(A)\subset \mbb{D}$. We hope that $\sigma(A)$ intersects with the complement of $\mbb{D}$, and in fact, contains the out-most eigenvalue of $F=\mr{D}f(0)$. 
Indeed, if the origin is a \emph{hyperbolic} equilibrium point, then in a neighborhood $\mbb{X}_0$ of the origin, there is a homeomorphism $\psi$ that brings the system to $z_{t+1}=Fz_t$. Hence, by finding the eigenvector $v$ corresponding to the out-most eigenvalue $\lambda_1(F)$, a Koopman eigenfunction $\phi(x)=v^\top \psi(x)$ can be found, which belongs to $\mc{H}$ as long as $\psi$ is a $\Phi^s$-homeomorphism. 
\par To extend the definition of the eigenfunction from $\mbb{X}_0$ to the entire $\mbb{X}$, we leverage the idea of \emph{open eigenfunctions} \citep{korda2020optimal, mezic2020spectrum}. That is, we backtrack all points $x\in \mbb{X}$ to $\mbb{X}_0$ and define for all $x$:
\begin{equation}\label{eq:open.eigenfunction}
\textstyle
    \phi(x) = \lambda_1(F)^{\tau(x)} v^\top \psi\bra{ f^{-\tau(x)}(x)} \text{ where } \tau(x) = \inf\{\tau\in \mbb{N}: f^{-\tau}(x)\in \mbb{X}_0\}. 
\end{equation}
The following theorem states that if all points in $\mbb{X}$ is reachable by an ``outbound'' trajectory issued from $\mbb{X}_0$ within a uniform time $\overline{\tau}<\infty$, then the function constructed in \eqref{eq:open.eigenfunction} is meaningful. 
Again, the homeomorphism of Theorem \ref{th:bifurcation} roughly requires that the equilibrium point at $0$ is the sole invariant structure in $\mbb{X}$ considered. 

\begin{theorem}\label{th:bifurcation}
    Assume the conditions in Theorem \ref{th:definedness} and the existence of a $\Phi^s$-homeomorphism $\psi$ on a compact $\mbb{X}_0 \subset \mbb{X}$ such that $\tilde{f} = \psi\circ f\circ \psi^{-1}: z\mapsto Fz$. Suppose that $\sup_{x\in \mbb{X}} \tau(x) = \overline{\tau} < \infty$ and furthermore $f^{-1}\in (\Phi^s)^d$. Then $\phi$ defined in \eqref{eq:open.eigenfunction} belongs to $\mc{H}$ and is an eigenfunction associated with eigenvalue $\lambda_1(F)$. 
\end{theorem}

\subsection{Learned Koopman operator for stability analysis} 
After all, when the Koopman operator is learned from data (i.e., an i.i.d. sample $\{(x^{(i)}, y^{(i)})\}_{i=1}^n$), there is an error contained in the estimation $\hat{A}$. 
Since $\hat{A}$ is finite-rank (and hence compact) while $A$ may not be compact from $\mc{H}$ to $\mc{H}$, it is generally impossible to bound the error on $\mc{H}$. Nevertheless, $\mc{H}$ can be compactly embedded into $L^2(\mbb{X})$ when $\mbb{X}$ is a compact region of $\mR^d$, and hence it is possible to bound $\|\hat{A} - A\|_{\mc{H}\to L^2(\mbb{X})}$. The following sharp bound is given in \cite{kostic2023sharp}.   
\begin{lemma}
    Suppose that the conditions in Theorem \ref{th:definedness} hold, and the eigenvalues of the covariance operator $\int_{\mbb{X}}\kappa(x,\cdot)\times \kappa(x,\cdot) \xD{x}$ on $L^2(\mbb{X})$: $\mu_1>\mu_2>\dots$ such that $\mu_j\leq bj^{-1/\alpha}$ for some $\alpha\in (0, 1]$ and $b>0$. Let $\hat{A}$ be the optimal reduced rank estimation in \eqref{eq:kEDMD}, where the points are sampled on $\mbb{X}$ uniformly. Then it holds with probability at least $1-\delta$ that
    \begin{equation}
        \|\hat{A}-A\|_{\mc{H}\rightarrow L^2(\mbb{X})} \leq \sigma_{r+1}(A) + \frac{n^{-1/(1+\alpha)} \log(1/\delta)}{\sigma_r^2(A)-\sigma_{r+1}^2(A)}. 
    \end{equation} 
\end{lemma}
Here $\sigma_{r+1}(A)$ refers to the $(r+1)$-th largest singular value of $A$. When varying the rank $r$ in the estimation, the decreasing $\sigma_{r+1}$ and the increasing $1/(\sigma_r^2-\sigma_{r+1}^2)$ requires a tradeoff. Such a tradeoff depends on the desired $\epsilon$ and hence its required sample size $n$. Specifically, larger dataset allows one to use a higher rank $r$ to achieve lower generalization error. In simple terms, we say that $\|\hat{A}-A\|_{\mc{H}\rightarrow L^2(\mbb{X})} = \epsilon_{n,\delta}$ with high probability (at least $1-\delta$), and that $\epsilon_{n,\delta}\rightarrow 0$ as $n\rightarrow \infty$. 

\par We then must consider the problem of applying this perturbation in $\mc{H}\rightarrow L^2(\mbb{X})$ to the spectrum. The following lemma from the perturbation theory of \emph{self-adjoint} linear operators on Hilbert spaces \citep{lax2002functional} shows that the error in the operator norm implies an error in the spectrum. 
In our context, we let $T = A^\ast A$, which then is a linear bounded operator on $\mc{H}$. 
\begin{lemma}\label{lem:perturbation}
    Suppose $T: \mc{H}\rightarrow\mc{H}$ is self-adjoint and $\sigma(T)$ is discrete. Let $E_T$ be a self-adjoint bounded linear operator on $\mc{H}$. Then $\sigma(T+E_T) \subset \{\lambda + \epsilon: \lambda\in \sigma(T), |\epsilon|\leq \|E_T\|\}$. 
\end{lemma}

Comparing $\hat{T} = \hat{A}^\ast \hat{A}$ and $T = A^\ast A$, we have $\|T-\hat{T}\| \leq  \|\hat{A}-A\|^2 + 2\|A\|\cdot \|A-\hat{A}\| = (\epsilon_{n,\delta} + 2\|A\|)\epsilon_{n,\delta}$. Hence, the spectral radius of $\hat{T}$ does not exceed that of $T$ added to $(\epsilon_{n,\delta} + 2\|A\|)\epsilon_{n,\delta}$. 
Here, we need to guarantee a discrete spectrum of $T= A^\ast A$, for which it suffices to make $A$ a diagonalizable operator. Under the homeomorphism argument of Theorem \ref{th:spectrum}, we only requires that the eigenvalues in $\overline{\<\sigma(F)\>}$ are \emph{disjoint}. The proof is given in the Appendix \ref{app:proofs}. 
\begin{lemma}\label{lem:self.adjoint}
    Assume that the conditions of Theorem \ref{th:spectrum} hold. If $\overline{\<\sigma(F)\>}$ does not have repeated entries (i.e., one cannot obtain the same value by multiplying $\lambda_1(F), \dots, \lambda_d(F)$ through different multi-indices), then $A^\ast A$ has a discrete spectrum, with a spectral radius of $|\lambda_1(F)|^2$, if $\sigma(F)\subset \mbb{D}$. 
\end{lemma}

Finally, we have the following theorem as a natural implication of Lemmas \ref{lem:perturbation} and \ref{lem:self.adjoint}. This allows us to certify the stability of the equilibrium point based on the learned Koopman operator $\hat{A}$. 
\begin{theorem}\label{th:final}
    Assume that the conditions of Theorem \ref{th:spectrum} and Lemma \ref{lem:self.adjoint} hold. If the spectrum of the learned Koopman operator $\hat{A}$ from kernel EDMD \eqref{eq:kEDMD} satisfies $\sigma_1(\hat{A})^2 + (\epsilon_{n,\delta}+2\|A\|)\epsilon_{n,\delta} < 1$, then with probability at least $1-\delta$, the equilibrium point is asymptotically stable. 
\end{theorem}

\section{Numerical Example}\label{sec:numerical}
We evaluate the proposed methodology numerically on the Van der Pol oscillator:
\begin{equation}\label{eq_system_VanDerPol}
\nonumber
    \dot{x}_1 = x_2, \qquad 
    \dot{x}_2 = \mu(1 - x_1^2)x_2 - x_1,
\end{equation}
with a parameter $\mu \in \mathbb{R}$. Two qualitatively distinct parameter regimes are considered: (i) for $\mu < 0$, the origin is a globally asymptotically stable equilibrium; (ii) for $\mu > 0$, the origin is unstable and the trajectories approach a limit cycle.
The dynamics is assumed unknown while data samples are available. Specifically, from $40$ randomly initialized trajectories with a sampling interval $0.2$, we learn a finite-rank Koopman approximation via \eqref{eq:kEDMD} using our linear--radial product kernel.
 \begin{figure}[t]
    \centering
    % --- Stable case ---
    \begin{minipage}[b]{0.455\columnwidth}
        \centering
        \includegraphics[width=1.2\linewidth]{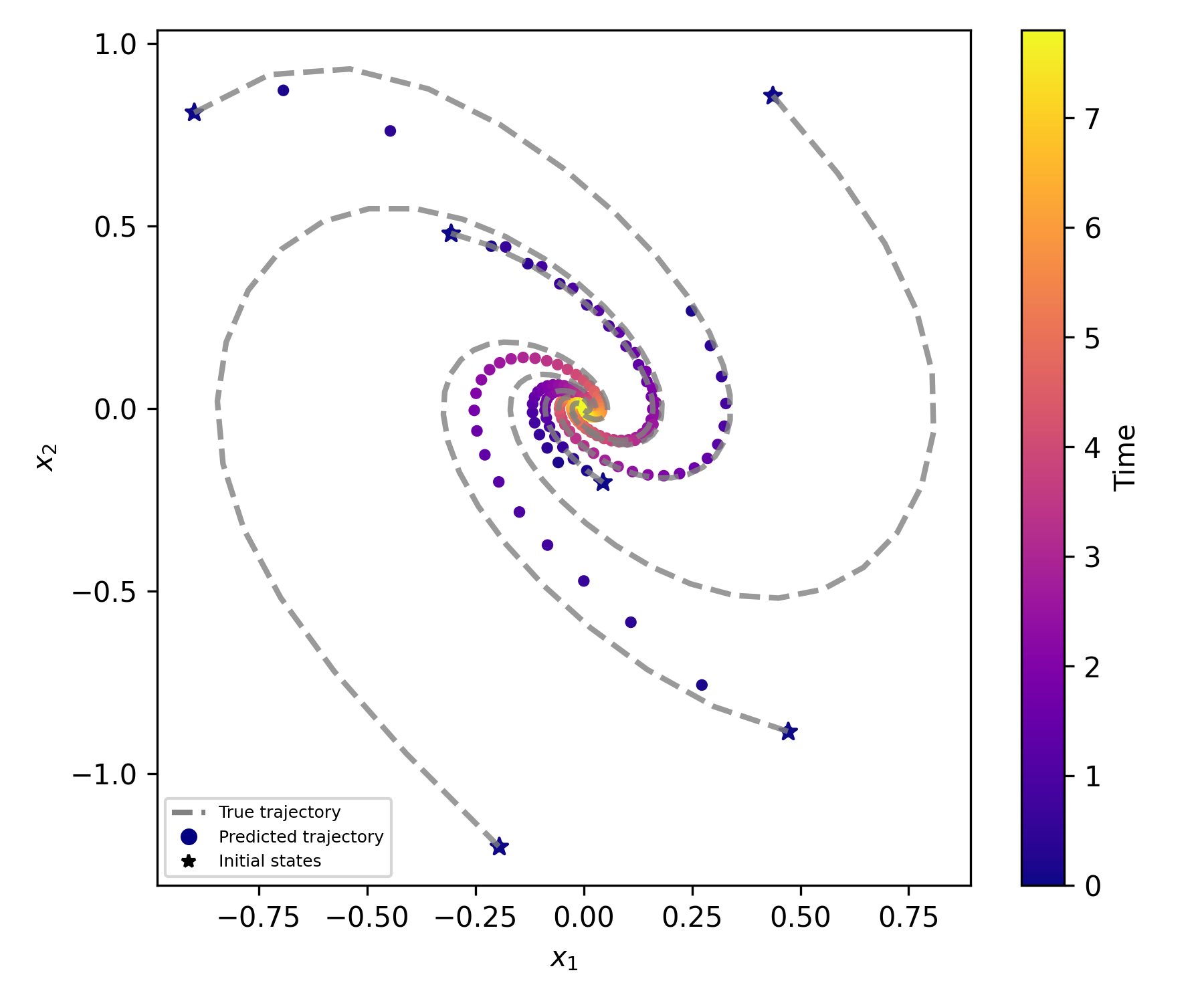}
        \caption{Trajectory prediction based on estimated Koopman operator when $\mu=-1$.}
        \label{fig1}
    \end{minipage}
    \hfill
    \begin{minipage}[b]{0.455\columnwidth}
        \centering
        \includegraphics[width=1\linewidth]{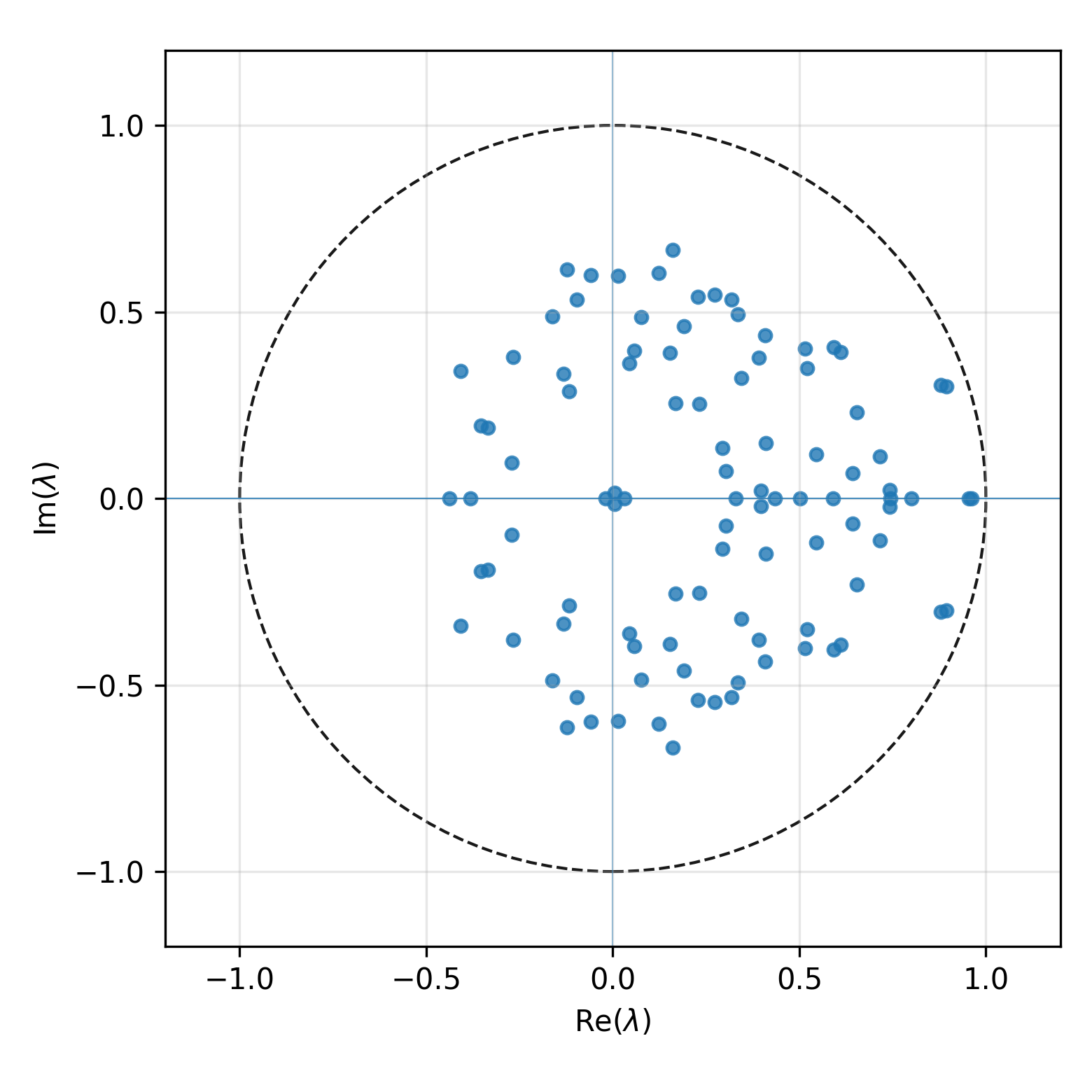}
        \caption{Spectrum of the estimated Koopman operator when $\mu=-1$.}
        \label{fig2}
    \end{minipage}
    \begin{minipage}[b]{0.455\columnwidth}
        \centering
        \includegraphics[width=1.2\linewidth]{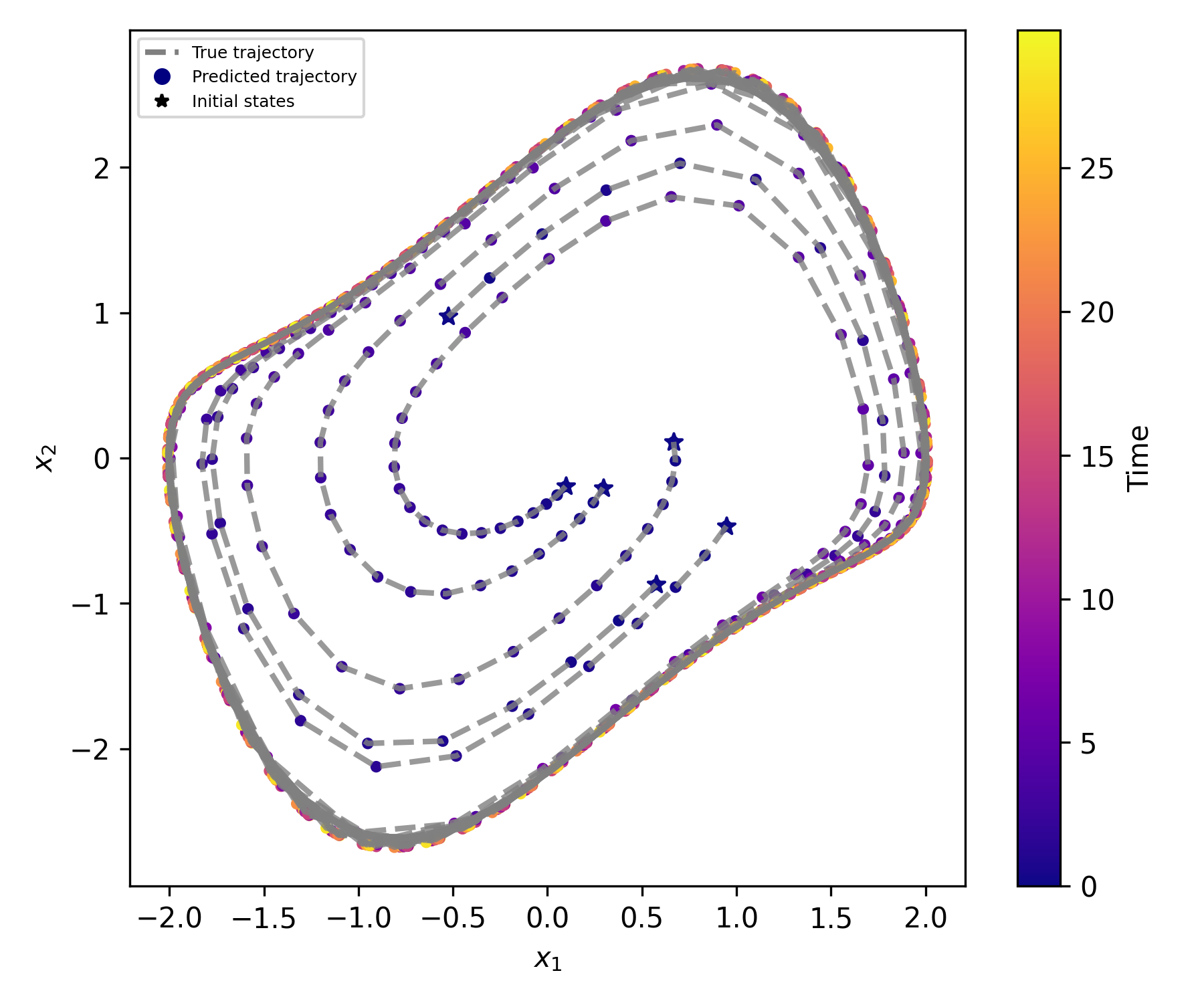}
        \caption{Trajectory prediction based on estimated Koopman operator when $\mu=+1$.}
        \label{fig3}
    \end{minipage}
    \hfill
    \begin{minipage}[b]{0.455\columnwidth}
        \centering
        \includegraphics[width=1\linewidth]{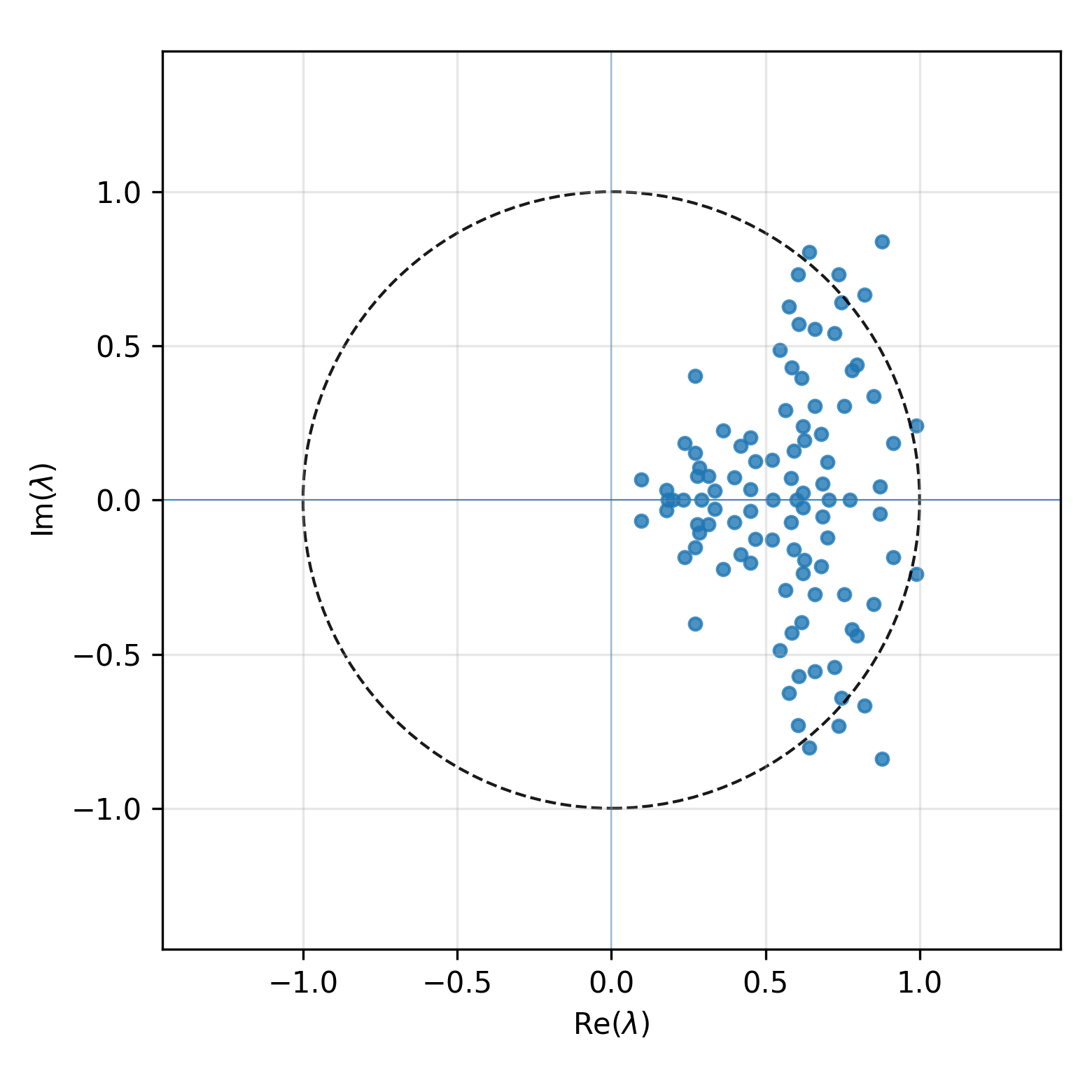}
        \caption{Spectrum of the estimated Koopman operator when $\mu=+1$.}
        \label{fig4}
    \end{minipage}
    \vspace{-1.0em}
\end{figure}

Fig.~\ref{fig1} shows the predicted trajectories compared to ground-truth trajectories when $\mu = -1$. The predictions resemble the true orbits and converge to the origin in a short time, which indicates that the estimated Koopman operator successfully captures the attraction toward the origin. 
This behavior is consistent with the spectrum of the learned Koopman operator in Fig.~\ref{fig2}, which lies strictly inside the unit circle. 
On the other hand, due to the effect of regularization and rank restriction in \eqref{eq:kEDMD}, the learned operator possesses many zero eigenmodes as an artifact, causing the predicted trajectories to converge faster to the origin than the actual dynamics. 

When $\mu = +1$, we evaluate the trajectory prediction shown in Fig.~\ref{fig3} with corresponding Koopman spectrum in Fig.~\ref{fig4}. We observe that the predicted trajectories follow tightly with the ground truth trajectories, moving away from the origin and falling onto the limit cycle. 
The fact that the equilibrium point is unstable is verified by the observation that the spectrum points of $\hat{A}$ overflows the unit circle. This confirms the conclusion of Theorem~\ref{th:bifurcation}. 
Here, we must sample the initial states cautiously close to the origin, so that the conditions of Theorem~\ref{th:bifurcation} are satisfied; that is, the entire $\mbb{X}$ is reachable from a small neighborhood $\mbb{X}_0$ of the origin and $\mbb{X}$ does not contain another attractor (the limit cycle) inside. (See Appendix \ref{app:simulation} for the results when sampling from a large domain. When $\mbb{X}$ contains the limit cycle, the spectrum learned is inconclusive about the stability of origin.)

\section{Discussions on Implications on Koopman-based Control}\label{sec:discussions}
\par Based on the idea that $\sigma(A)\subset \mbb{D}$ when the equilibrium point is asymptotically stable, we hypothesize that a Lyapunov function can be found by the Koopman operator $A$ on our new RKHS (and hence can be estimated through the learned version $\hat{A}$). 
Specifically, in an analogous manner to finite-dimensional linear systems, a Lyapunov function may be of a ``\emph{kernel quadratic form}'' $V(x) = \ip{\kappa(x, \cdot)}{P\kappa(x, \cdot)}$ that satisfies a ``\emph{kernel Lyapunov equation}'',
\begin{equation}\label{eq:Lyapunov} 
    V(f(x))-V(x)=\ip{\kappa(x, \cdot)}{(APA^\ast - P)\kappa(x, \cdot)} = \ip{\kappa(x, \cdot)}{-Q\kappa(x, \cdot)} =: -w(x).
\end{equation}
Here the operator $Q$ specifying the decay rate function $w$ is a Hilbert--Schmidt operator, which can be written as $Q=\sum_{i=1}^\infty \mu_i g_i\times g_i$ for an orthonormal basis $\{g_i\}_{i=1}^\infty$ in $\mc{H}$ with $\{\mu_i\}_{i=1}^\infty\downarrow 0$, i.e., $w(x)= \sum_{i=1}^\infty \mu_ig_i(x)^2$. 
Since $g_i = \sum_{j=1}^d e_j\phi_{ij}$ for some $\phi_{ij}\in W^{s,2}(\mbb{X})$ (by Lemma \ref{lem:member}), we see that $w = \sum_{i=1}^\infty \sum_{j=1}^d \mu_i(e_j\phi_{ij})^2$ is a ``locally at least quadratic function''. 
Now that $\sigma(A)\subset \mbb{D}$, we should be able to express the solution to the kernel Lyapunov equation \eqref{eq:Lyapunov} by $P = \sum_{t=0}^\infty A^tQA^{*t}$, thus obtaining a ``locally at least quadratic but globally generic'' Lyapunov function. The extension to continuous-time systems, using the notion of Koopman semigroup, is natural.

\par As in linear systems theory \citep{antsaklis2006linear, hespanha2018linear}, the controller synthesis can be formulated. For a continuous-time input-affine system $\dot{x} = f_0(x) +\sum_{j=1}^m u_jf_j(x)$, let $A_j: g\mapsto \nabla g\cdot f_j$ ($0\leq j\leq m$) be the infinitesimal generators of \emph{open-loop Koopman semigroups}. The feedback controller $u_j = \pi_j(x)$ results in a \emph{closed-loop Koopman generator} of $A=A_0+\sum_{j=1}^m M_{\pi_j}A_j$ where $M_{\pi_j}: g\mapsto g\cdot \pi_j$ are multiplication operators. 
The determination of a stabilizing control law thus reduces to finding $\pi_j$ satisfying a kernel Lyapunov equation:
\begin{equation}\label{eq:control.Lyapunov} 
\textstyle
    \< \kappa(x, \cdot), \Bra{ P\bra{A_0+\sum_{j=1}^m M_{\pi_j}A_j}^\ast + \bra{A_0+\sum_{j=1}^m M_{\pi_j}A_j}P } \kappa(x, \cdot) \> = \< \kappa(x, \cdot), -Q\kappa(x, \cdot)\>.
\end{equation}
A rigorous theory on the solution existence, uniqueness, and learning error is left for future effort.

\section{Conclusion}\label{sec:Conclusion}
In this paper, we have proposed a novel RKHS space, specified by linear--radial product kernels, to define the Koopman operator for discrete-time nonlinear systems. 
With only mild assumptions on a known equilibrium point and smoothness, the Koopman operator is a bounded linear operator, which can be learned from a (large but finite) data sample with a guaranteed error bound.  
Most importantly, we have shown that under appropriate homeomorphism assumptions, the asymptotic stability of equilibrium point is reflected by a Koopman spectrum confined in the open unit disk, while upon bifurcation, the spectral radius exceeds $1$. As such, \emph{the learned Koopman operator's spectrum provides a stability certificate}. 

\par We underscore that such a certification did not exist with existing RKHS constructions for the Koopman operator. 
Due to the our established correspondence between Koopman spectrum and stability of the equilibrium point, we have discussed, conceptually, the potential use of this novel RKHS framework for Koopman-based \emph{direct synthesis of stabilizing controllers}. % This is undergoing active investigation by the authors. 
Another promising direction is to account for other forms of invariant sets other than equilibrium points, such as an attractive/repulsive limit cycle. It would also be of interest to examine whether a properly defined Koopman operator can capture the behaviors of multiple invariant structures simultaneously. 

\newpage
\acks{This work is supported by National Science Foundation under CBET Award \#2414369 and the faculty startup fund from NC State University. The codes for numerical experiments, including the supplementary ones in the Appendix, are available at \href{https://github.com/XiuzhenYe/Koopman-Operator-for-Stability-Analysis-Theory-with-a-Linear-Radial-Product-Reproducing-Kernel}{this GitHub repository}. 
The authors would like to thank Petar Bevanda for bringing to our attention the works on the statistical error analysis of Koopman operator learning and providing clarifications on pertinent ideas. 
} % The authors would like to thank Prof. I. G. Kevrekidis at Johns Hopkins University and Prof. Jun Liu at University of Waterloo for beneficial discussions and encouragement.
\bibliography{mybib}

\newpage
\begin{center}
    \textbf{\Large \textit{Supplementary Information to} ``Koopman Operator for Stability Analysis: Theory with a Linear--Radial Product Reproducing Kernel''} \\[1.5em]
    \textbf{Wentao Tang} and \textbf{Xiuzhen Ye} \\ 
    \textit{Department of Chemical and Biomolecular Engineering, North Carolina State University}
\end{center}

\appendix
\section{Technical Proofs Omitted in the Main Text}\label{app:proofs}
\subsection{Proof of Lemma \ref{lem:member}}
\begin{proof}
    The representation \eqref{eq:member} is obvious by  definition. For $g = \sum_{i=1}^d e_ih_i \in \mc{H}$, 
    $$\|g\|^2 = \sum_{i=1}^d \sum_{j=1}^d \ip{e_i\otimes h_i}{e_j\otimes h_j} = \sum_{i=1}^d \sum_{j=1}^d \ip{e_i}{e_j}_{H_{\kappa^-}} \ip{h_i}{h_j}_{H_{\kappa^=}},$$
    where $\ip{e_i}{e_j}_{H_{\kappa^-}} = \kappa^-(\mb{e}_i, \mb{e}_j) = \delta_{ij}$. By $\mb{e}_i$ we refer to the $i$-th standard orthonormal basis vector of $\mR^d$, and $\delta_{ij}$ the Kronecker's delta. The conclusion is hence made apparent.
\end{proof}

\vspace{-1em}
\subsection{Proof of Theorem \ref{th:definedness}}
\begin{proof}
    By condition (ii), for all $1\leq i\leq d$, we have $f_i = \sum_{j=1}^d e_j\phi_{ij}$, in which $\phi_{ij}\in C^s(\mbb{X})$ for all $1\leq j\leq d$. Then for any $g= \sum_{i=1}^d e_ih_i$ where $h_i\in W^{s,2}(\mbb{X})$ for $1\leq i\leq d$, we obtain $Ag = \sum_{i=1}^d (e_i\circ f)(Ah_i) = \sum_{i=1}^d f_i(Ah_i) = \sum_{i=1}^d\sum_{j=1}^d e_j\phi_{ij}(Ah_i)$. By Lemma \ref{lem:member}, we have
    $$\|Ag\|^2 = \sum_{j=1}^d \norm{\sum_{i=1}^n \phi_{ij}Ah_i}_{=}^2 
    \leq \mr{const}\cdot \Bra{\sum_{i=1}^d\sum_{j=1}^d \norm{\phi_{ij}}_{C^s}^2} \cdot \Bra{\sum_{i=1}^n \norm{Ah_i}_{H_{\kappa^=}}^2} $$
    in which the $C^s$-norm of any $\phi\in C^s(\mbb{X})$ can be defined by $\|\phi\|_{C^s}^2 = \sum_{|\alpha|\leq s} \sup_{x\in \mbb{X}} |\partial^\alpha \phi(x)|^2$ ($\alpha$: multi-indices). Since $A$ on $H_{\kappa^=} \simeq W^{s,2}$ is bounded by Lemma \ref{lem:smoothness} under conditions (i) and (iii), $$ \|Ag\|^2 \leq \mr{const}\cdot \Bra{\sum_{i=1}^d \norm{h_i}_{H_{\kappa^=}}^2} = \mr{const}\cdot \|g\|^2 $$
    must hold, where we reused Lemma \ref{lem:member}. This completes the proof. 
\end{proof}

\vspace{-1em}
\subsection{Proof of Lemma \ref{lem:homeomorphism}}
\begin{proof}
    Since the components of $\psi$ and $\psi^{-1}$ belong to the $\Phi^s$-class, analogous to the proof in Theorem \ref{th:definedness}, the composition operators $C_\psi: \mc{H}\rightarrow \tilde{\mc{H}}$, $g\mapsto \psi\circ g$ and $C_{\psi^{-1}}: \tilde{\mc{H}} \rightarrow \mc{H}$, $\tilde{g}\mapsto \psi^{-1}\circ \tilde{g}$ are both linear and bounded. Hence $\tilde{A} = C_{\psi}AC_{\psi^{-1}}$ is linear and bounded. 
    Clearly, for any $\lambda\in \mC$, $\lambda I - A$ has a bounded inverse on $\mc{H}$ if and only if $\lambda I - \tilde{A}$ has a bounded inverse on $\tilde{H}$. Hence, their sets of regularity points, and hence the sets of spectrum points, coincide. 
\end{proof}

\vspace{-1em}
\subsection{Proof of Theorem \ref{th:spectrum}}
\begin{proof}
    Without loss of generality assume that $\{\ms{e}_i\}_{i=1}^d$ are the eigenvectors and generalized eigenvectors determining a Jordan canonical form of $F$. Thus, the linear functions $\{e_i\}_{i=1}^d$ are the eigenfunctions and generalized eigenfunctions of $A$, and the eigenvalues are $\sigma(F)$. Hence, all monomials of $x$ of degree $r$, e.g., $x_1^{\alpha_1}\cdots x_d^{\alpha_d}$, is a generalized eigenfunction of $A$, with eigenvalue of the form $\lambda_1^{\alpha_1} \cdots \lambda_d^{\alpha_d}$. Therefore, for all $\lambda \notin \overline{\< \sigma(F)\>}$, $(\lambda I-A)^{-1}$ exists and is in fact a bounded linear operator on the space spanned by the monomials, namely the space of polynomials. 
    \par For a compact region $\mbb{X}$ with a Lipschitz boundary, the space of polynomials in dense in $W^{s,2}(\mbb{X})$ \citep{adams2003sobolev}. It then follows that the space of polynomials with zero constant terms is dense in $\mc{H}$, according to Lemma \ref{lem:member} on the representation of function members of $\mc{H}$. Therefore, by the previous paragraph, for all $\lambda \notin \overline{\< \sigma(F)\>}$, $(\lambda I-A)^{-1}$ is a bounded linear operator on $\mc{H}$. Hence $\sigma(A)= \overline{\< \sigma(F) \>}$. Since $\sigma(F)$ contains at most $d$ numbers, if $F$ is stable, any $\lambda\in \sigma(F)$ satisfies $|\lambda|< 1$ and hence we have $\sup_{\lambda\in \overline{\< \sigma(F) \>}} |\lambda|< 1$. Then the proof is complete. 
\end{proof}

\vspace{-1em}
\subsection{Proof of Theorem \ref{th:bifurcation}}
\begin{proof}
    Let $\mbb{X}_1 = f(\mbb{X}_0) = \{f(x): x\in \mbb{X}_0\}$ and $\mbb{X}_{t} = f(\mbb{X}_{t-1})$ for $t=2,\dots,\overline{\tau}$ recursively. All these sets are compact. On each $\mbb{X}_t$, the definition of $\phi$ is given by $\phi(x)= \lambda^t A^{-t}\phi(x)$. 
    Under the supposed regularity for $f^{-1}$, $A^{-1}$ is a linear and bounded operator on $\mc{H}$. Clearly, $\phi|_{\mbb{X}_0}$ belongs to $\mc{H}$, since $\psi$ is a $\Phi^s$-homeomorphism. Hence, when restricted to $\cup_{t=0}^{\overline{\tau}} \mbb{X}_t$,  $\phi$ is well-defined and still belongs to $\mc{H}$. 
    The fact that its eigenvalue is $\lambda_1(F)$ is easily verified. 
\end{proof}

\vspace{-1em}
\subsection{Proof of Lemma \ref{lem:self.adjoint}}
\begin{proof}
    If $\overline{\<\sigma(F)\>}$ does not contain repeated elements, then all eigenvalues of $F$ are distinct, and hence $F$ is a diagonalizable matrix, whose eigenvectors form a basis of $\mR^d$. 
    It follows that there are also no generalized eigenfunctions corresponding to any eigenvalue of $A$, which implies that the Koopman operator $A$ is a diagonal operator on $\mc{H}$. As such, for any $h\in \mc{H}$ with $\|h\|_{\mc{H}} = 1$, we have $\|Ah\|^2 \leq \max_{\lambda\in \sigma(A)} |\lambda|^2$. 
    Since $\sigma(A) = \overline{\<\sigma(F)\>}$, if $\sigma(F)\subset \mbb{D}$, we have $\max_{\lambda\in \sigma(A)} |\lambda|^2 = |\lambda_1(F)|^2$. Thus, $\|Ah\|^2\leq |\lambda_1(F)|^2$, i.e., $\ip{h}{A^\ast Ah}\leq |\lambda_1(F)|^2$. Therefore, the spectral radius of $A$ is bounded by $|\lambda_1(F)|^2$. 
\end{proof}

\vspace{-1em}
\subsection{Proof of Theorem \ref{th:final}}
\begin{proof}
    Considering $T := A^\ast A$ as a self-adjoint compact operator from $\mc{H}$ to $\mc{H}$, its counterpart $\hat{T} := \hat A^\ast \hat A$ based on estimated Koopman operator $\hat A$ is a finite-rank operator. 
    Since $$\|\hat{T} - T\| \leq 2\|A\|_{\mc{H}\to L^2(\mbb{X})} \|\hat{A} - A\|_{\mc{H}\to L^2(\mbb{X})} + \|\hat{A} - A\|_{\mc{H}\to L^2(\mbb{X})}^2 , $$ 
    when $\|\hat{A} - A\|_{\mc{H}\to L^2(\mbb{X})}\leq \epsilon_{n,\delta}$, we have $\|\hat{T} - T\|\leq (2\|A\|+\epsilon_{n,\delta}) \epsilon_{n,\delta}$. By Lemma \ref{lem:perturbation}, 
    $$\sigma(T) \subset \BRA{ \lambda + \epsilon: \lambda\in \sigma(\hat{T}), |\epsilon|\leq (2\|A\|+\epsilon_{n,\delta}) \epsilon_{n,\delta} }, $$
    and hence the spectrum radii satisfy the inequality: $$\mr{radius}\, \sigma(T)\leq \mr{radius}\, \sigma(\hat T) + (2\|A\|+\epsilon_{n,\delta}) \epsilon_{n,\delta}.$$
    Since $\mr{radius}\, \sigma(\hat T) = \sigma_1(A)^2$ and $\mr{radius}\, \sigma(T) = |\lambda_1(A)|^2$ when $A$ is diagonal, the conclusion follows. 
\end{proof}

\section{Supplementary Numerical Simulations}\label{app:simulation} 
\subsection{Interference by Other Invariant Structures} 
\begin{figure}[t]
  \centering
  \begin{minipage}[b]{0.48\columnwidth}
    \centering
    \includegraphics[width=1.18\linewidth]{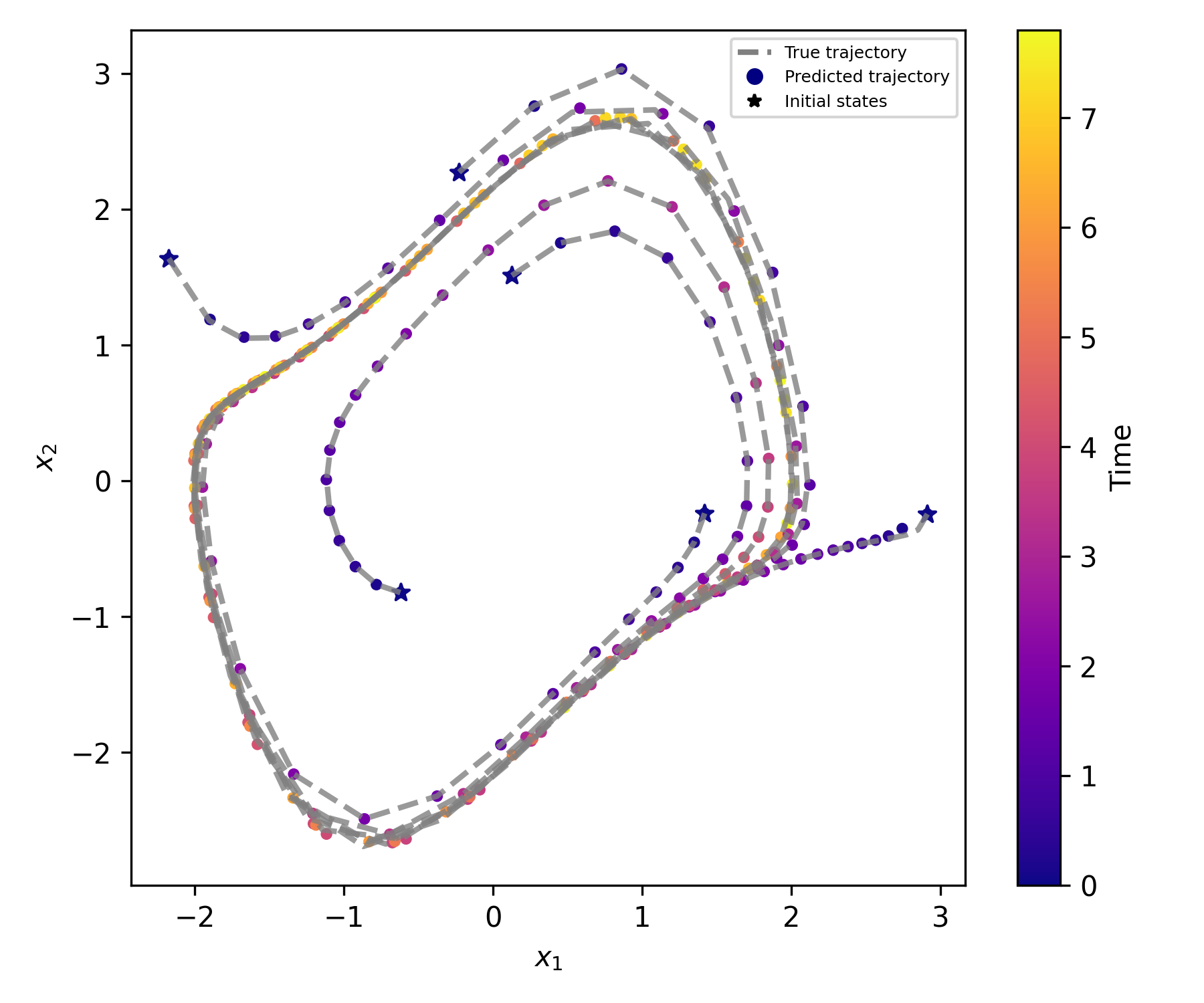}
    \caption{Trajectory prediction based on estimated Koopman operator when $\mu=+1$ with initial states in the domain $[-2.5, 2.5]$.}
    \label{fig5}
  \end{minipage}\hfill
  \begin{minipage}[b]{0.48\columnwidth}
    \centering
    \includegraphics[width=1\linewidth]{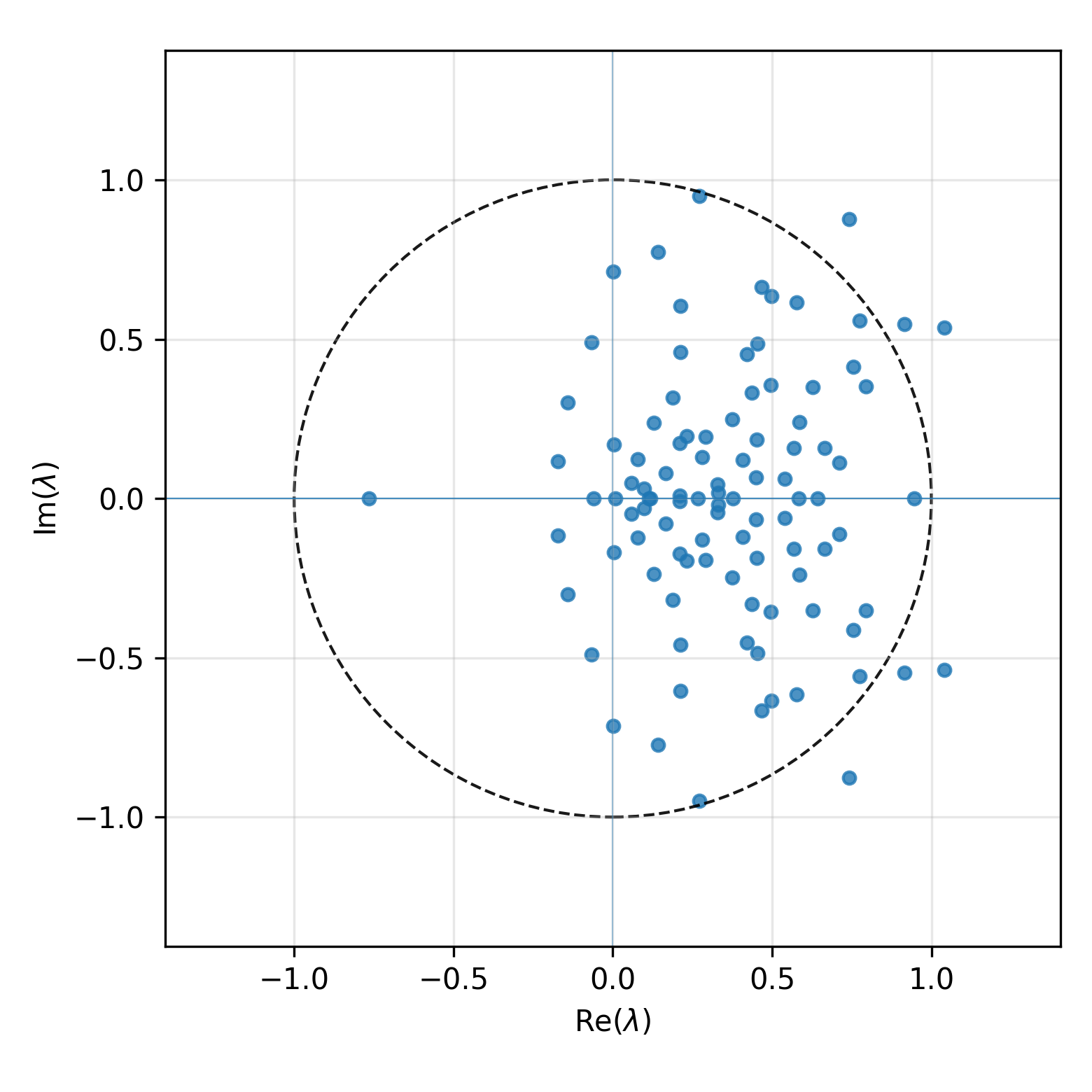}
    \caption{Spectrum of the estimated Koopman operator when $\mu=+1$ with initial states in the domain $[-2.5, 2.5]$.}
    \label{fig6}
  \end{minipage}
\end{figure}

In Section \ref{sec:numerical}, we verified that for the learned Koopman operator, if the initial states are sampled in a well-specified $\mbb{X}$ that is small enough to not contain any other invariant structure, then the Koopman spectrum is loyal to the stability property of the equilibrium point. 
For the case of $\mu=+1$ (when the origin is unstable), we also evaluated the performance of the proposed approach with initial states randomly sampled from a larger domain $[-2.5, +2.5]$, which contains an attractive limit cycle inside. 
The trajectory prediction based on the estimated Koopman operator is depicted in Fig.~\ref{fig5} with the Koopman spectrum shown in Fig.~\ref{fig6}. The estimated Koopman operator still predicts the trajectory very well empirically. 

\par In this case, the spectrum still shows an overflow from the unit disk. This indicates that from the collected data sample, it is possible to construct an approximate eigenfunction of the Koopman operator with an eigenvalue $|\lambda|>1$. This is a reflection of the existence of an attractive limit cycle. 

\vspace{-1em}
\subsection{Empirical Observations under Radial Kernel only} 
\par The proposed method in this paper is to use a linear--radial product kernel. The linear kernel serves as a special treatment of the equilibrium point, while the radial kernel (Wendland kernel) $\kappa^=$ plays the role of ensuring invariance of the RKHS under the action of Koopman operator. 
Theoretically, if defining $A: H_{\kappa^=} \rightarrow H_{\kappa^=}$, then it must hold $A^\ast \kappa^=(x, \cdot) = \kappa^=(f(x), \cdot)$. Since for any $t$, 
$$\|\kappa^=(x, \cdot)\|_{H_{\kappa^=}}^2 = \kappa^=(x,x) = \rho(0)= \kappa^=(f^t(x), f^t(x)) = \|\kappa^=(f^t(x), \cdot)\|_{H_{\kappa^=}}^2, $$
it is guaranteed that $\|A^t\|\geq 1$ and hence the spectral radius, by Gelfand formula, $\mr{radius}~\sigma(A) = \limsup_{t\rightarrow\infty} \|A^t\|^{1/t} \geq 1$, even if the origin is an asymptotically stable equilibrium point. In fact, in such a case, $\kappa^=(0, \cdot)$ is always an eigenfunction with eigenvalue $1$. 

\begin{figure}[!t]
\centering
% --- Stable case ---
\begin{minipage}[b]{0.455\columnwidth}
    \centering
    \includegraphics[width=1.2\linewidth]{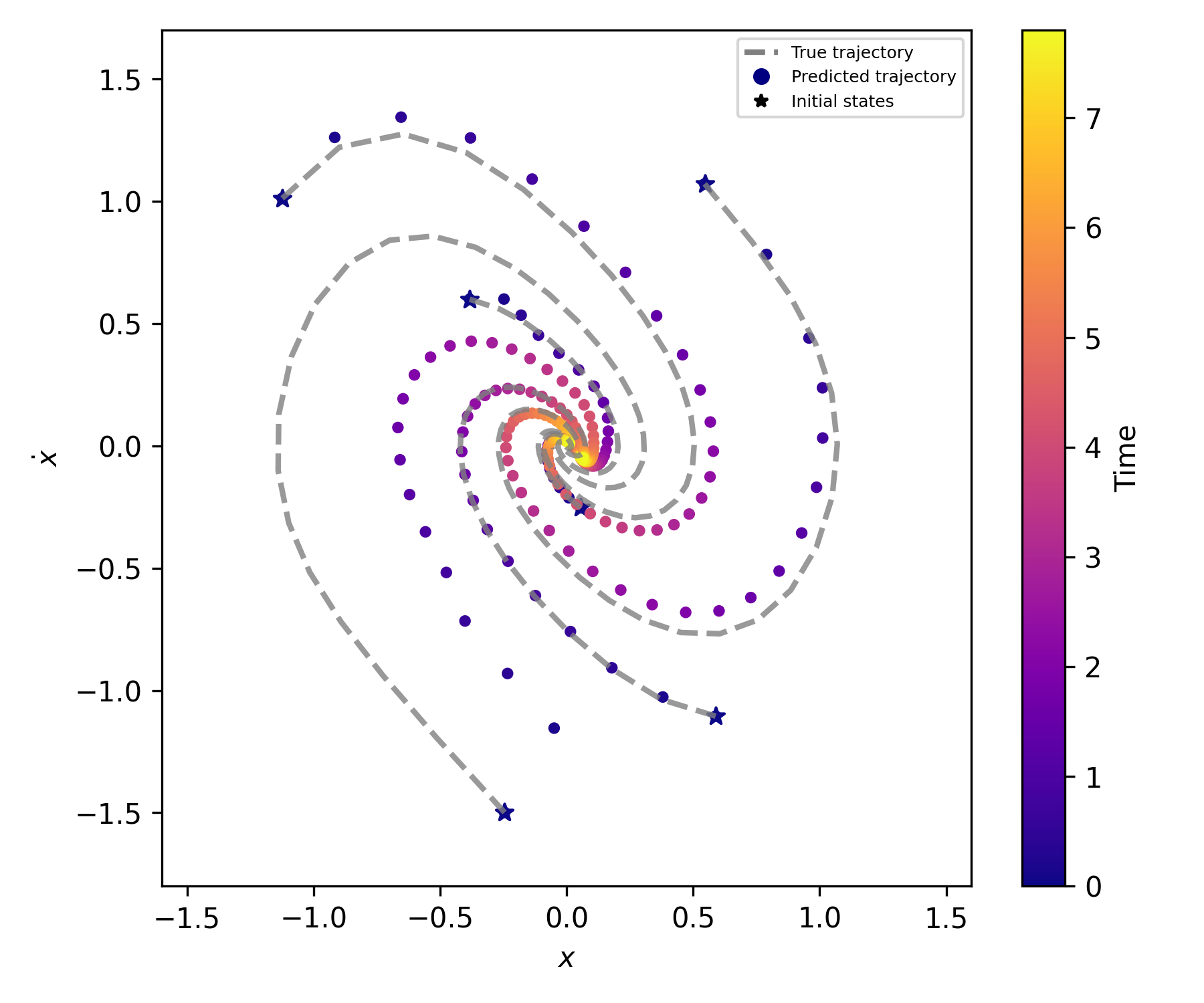}
    \caption{Trajectory prediction based on estimated Koopman operator using Wendland kernel only when $\mu=-1$.}
    \label{fig7}
\end{minipage}
\hfill
\begin{minipage}[b]{0.455\columnwidth}
    \centering
    \includegraphics[width=1\linewidth]{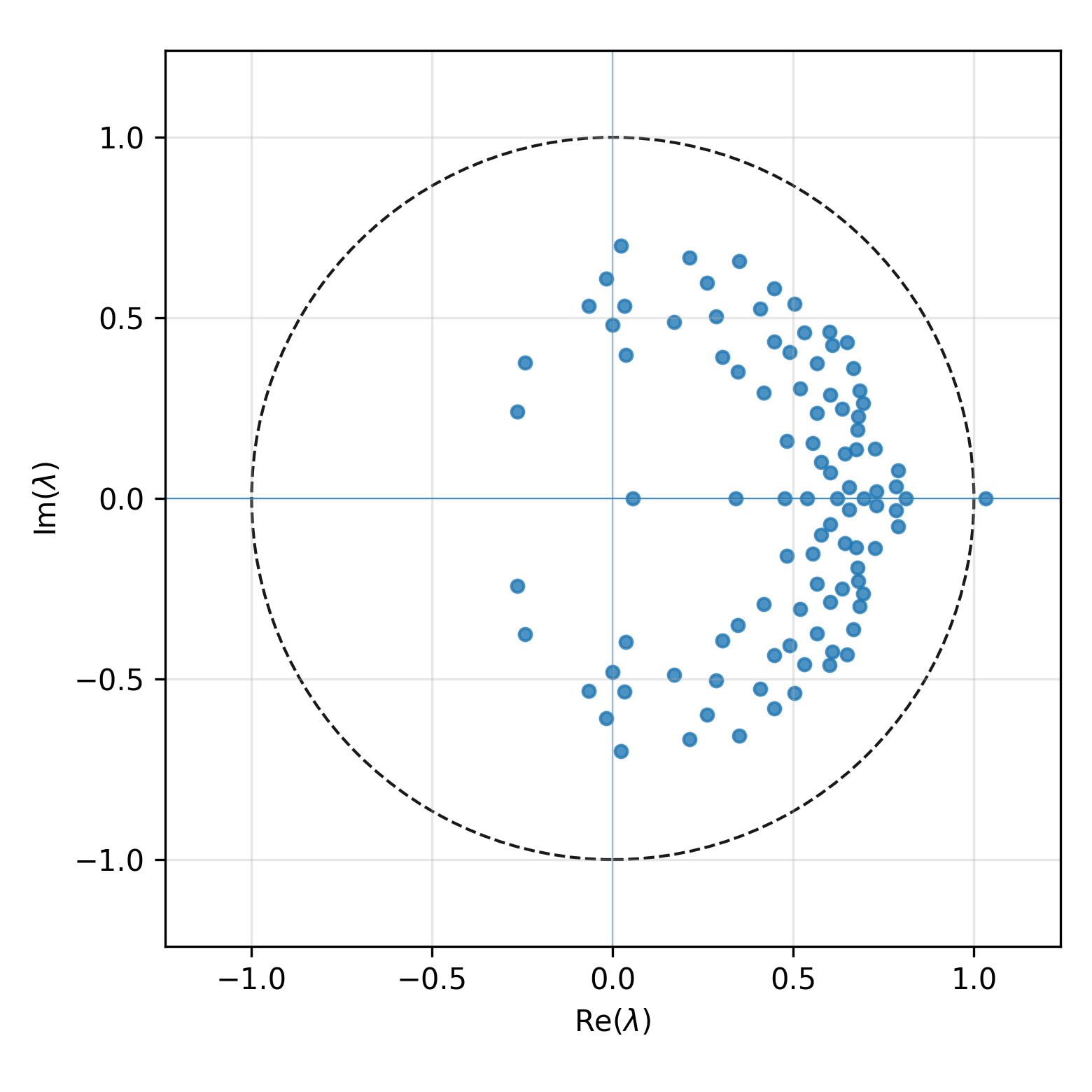}
    \caption{Spectrum of the estimated Koopman operator using Wendland kernel only when $\mu=-1$.}
    \label{fig8}
\end{minipage}
\begin{minipage}[b]{0.455\columnwidth}
    \centering
    \includegraphics[width=1.2\linewidth]{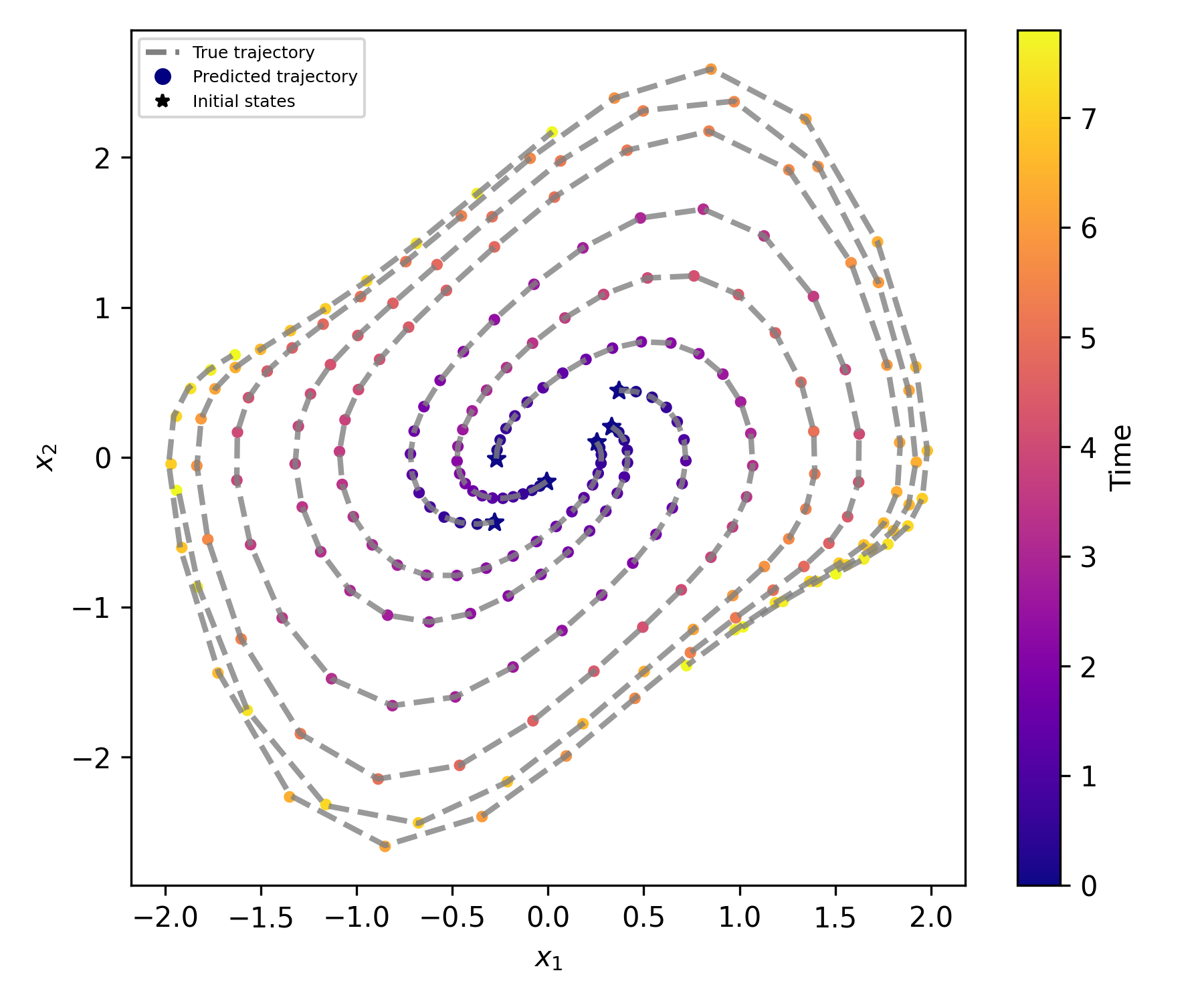}
    \caption{Trajectory prediction based on estimated Koopman operator using Wendland kernel only when $\mu=+1$.}
    \label{fig9}
\end{minipage}
\hfill
\begin{minipage}[b]{0.455\columnwidth}
    \centering
    \includegraphics[width=1\linewidth]{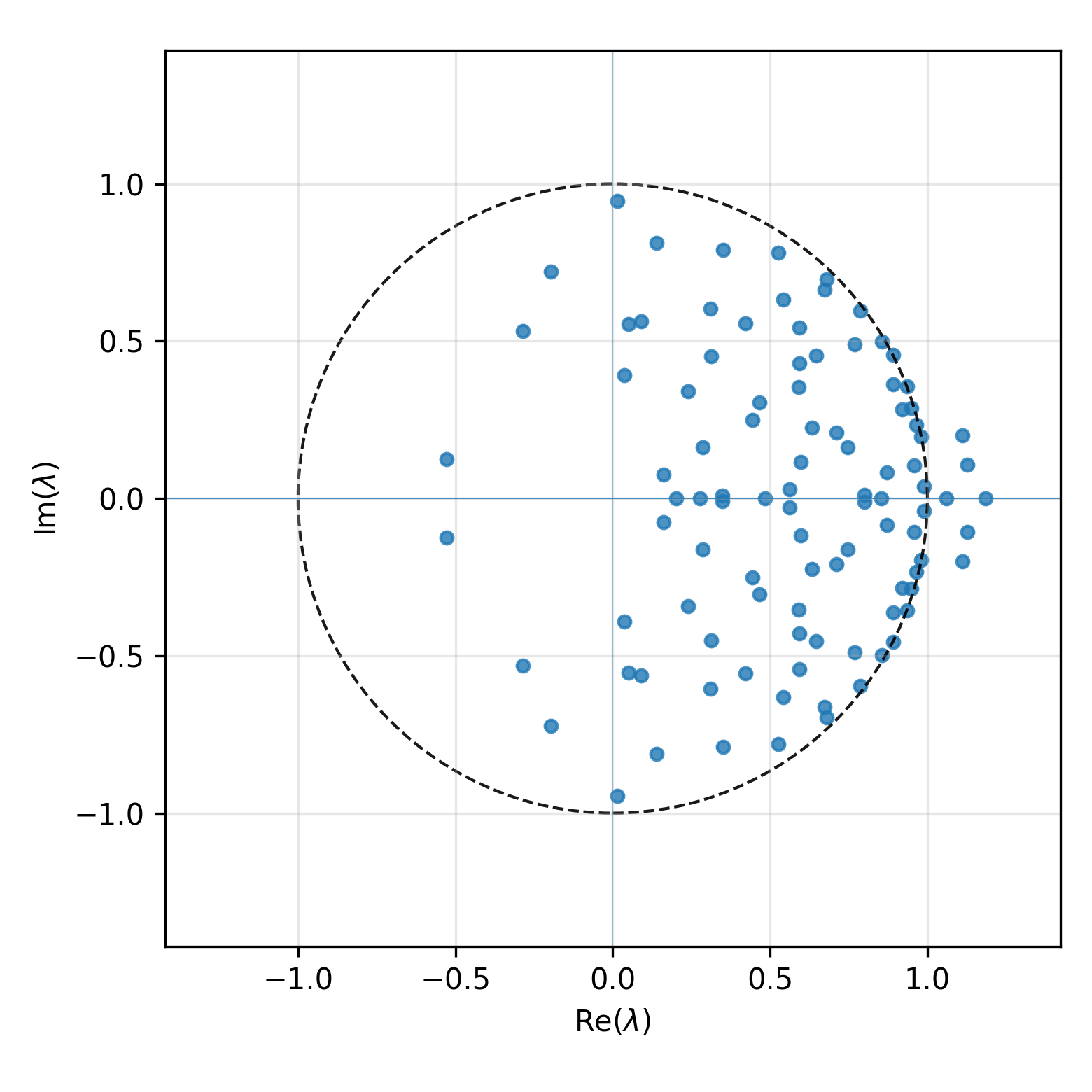}
    \caption{Spectrum of the estimated Koopman operator using Wendland kernel only when $\mu=+1$.}
    \label{fig10}
\end{minipage}
\vspace{-1.0em}
\end{figure}

\par The trajectory prediction and spectrum results, when using the Wendland kernel only, are shown in Figures \ref{fig7}--\ref{fig10}. We observed that the spectrum of $\hat{A}$ in both two cases of $\mu$ (Figures \ref{fig8} and \ref{fig10}) shows an overflow from $\mbb{D}$, confirming the shortage of spectrum--stability relation under a radial kernel. 

\end{document}